\documentstyle[12pt]{article}
\begin{document}

\title{{\bf Boundary Conditions and Predictions of Quantum Cosmology}
\thanks{Alberta-Thy-21-06, hep-th/0612194,
invited lecture for Session COT5 of the 11th Marcel Grossmann Meeting
on General Relativity, 2006 July 24}}
\author{
Don N. Page
\thanks{Internet address:
don@phys.ualberta.ca}
\\
Theoretical Physics Institute\\
Department of Physics, University of Alberta\\
Room 238 CEB, 11322 -- 89 Avenue\\
Edmonton, Alberta, Canada T6G 2G7
}
\date{(2006 December 15)}

\maketitle
\large
\baselineskip 18 pt

\begin{abstract}

	A complete model of the universe needs at least three parts:
(1) a complete set of physical variables and dynamical laws for them,
(2) the correct solution of the dynamical laws, and
(3) the connection with conscious experience.
In quantum cosmology, item (1) is often called a `theory of
everything,' and item (2) is the quantum state of the cosmos.  Hartle
and Hawking have made the `no-boundary' proposal, that the
wavefunction of the universe is given by a path integral over all
compact Euclidean 4-dimensional geometries and matter fields that
have the 3-dimensional argument of the wavefunction on their one and
only boundary.  This proposal has had several partial successes,
mainly when one takes the zero-loop approximation of summing over a
small number of complex extrema of the action.  However, it has also
been severely challenged by an argument by Susskind.

\end{abstract}

\normalsize

\baselineskip 14 pt

\newpage

\section{Introduction}\label{intro}

	A complete model of the universe needs at least three parts:

\begin{enumerate}

\item A complete set of physical variables (e.g., the arguments of
the wavefunction) and dynamical laws (e.g., the Schr\"{o}dinger
equation for the wavefunction, the algebra of operators in the
Hilbert space, or the action for a path integral.) Roughly speaking,
these dynamical laws tell how things change with time. Typically they
have the form of differential equations.

\item The correct solution of the dynamical laws (e.g., the
wavefunction of the universe). This picks out the actual quantum
state of the cosmos from the set of states that would obey the
dynamical laws. Typically a specification of the actual state would
involve initial and/or other boundary conditions for the dynamical
laws.

\item The connection with conscious experience (e.g., the laws of
psycho-physical experience) These might be of the form that tells
what conscious experience occurs for a possible quantum state for the
universe, and to what degree each such experience occurs (i.e., the
measure for each set of conscious experiences \cite{SQM}).

\end{enumerate}

	Item 1 alone is called by physicists a TOE or `theory of
everything,' but it is not complete by itself.  Even Items 1 and 2
alone are not complete, since by themselves they do not logically
determine what, if any, conscious experiences occur in a universe.

\section{The Hartle-Hawking Proposal for the Quantum State}

	Here I shall focus on Item 2, the quantum state of the
cosmos, and in particular focus on a proposal by Hawking \cite{H} and
by Hartle and Hawking \cite{HH} for this quantum state. They have
proposed that the quantum state of the universe, described in
canonical quantum gravity by what we now call the Hartle-Hawking
wavefunction, is given by a path integral over compact
four-dimensional Euclidean geometries and matter fields that each
have no boundary other than the three-dimensional geometry and matter
field configuration that is the argument of the wavefunction.

	In particular, the wavefunction for a three-geometry given by a
three-metric $g_{ij}(x^k)$, and for a matter field configuration
schematically denoted by $\phi^A(x^k)$, where the three-metric and the
matter field configuration are functions of the three spatial
coordinates $x^k$ (with lower-case Latin letters ranging over the three
values $\{1,2,3\}$), is given by the wavefunction
\begin{equation}
\psi[g_{ij}(x^k),\phi^A(x^k)]
= \int{\mathcal{D}}[g_{\mu\nu}(x^{\alpha})]
{\mathcal{D}}[\phi^{\Omega}(x^{\alpha})]
e^{-I[g_{\mu\nu},\phi^{\Omega}]},
\end{equation}
where the path integral is over all compact Euclidean
four-dimensional geometries that have the three-dimensional
configuration $[g_{ij}(x^k),\phi^A(x^k)]$ on their one and only
boundary.  Here a four-geometry are given by a four-metric
$g_{\mu\nu}(x^{\alpha})$, and four-dimensional matter field histories
are schematically denoted by $\phi^{\Omega}(x^{\alpha})$, both functions
of the four Euclidean spacetime coordinates $x^{\alpha}$ (with
lower-case Greek letters ranging over the four values $\{0,1,2,3\}$).

\section{Incompleteness of the Hartle-Hawking Proposal}

	The Hartle-Hawking `one-boundary' proposal is incomplete in
various ways.  For example, in quantum general relativity, using the
Einstein-Hilbert-matter action, the path integral is ultraviolet
divergent and nonrenormalizable \cite{GS}. This nonrenormalizability
also occurs for quantum supergravity \cite{Des}. String/M theory gives
the hope of being a finite theory of quantum gravity (at least for each
term of a perturbation series, though the series itself is apparently
only an asymptotic series that is not convergent.)  However, in
string/M theory it is not clear what the class of paths should be in
the path integral that would be analogous to the path integral over
compact four-dimensional Euclidean geometries without extra boundaries
that the Hartle-Hawking proposal gives when general relativity is
quantized.

	Another way in which the Hartle-Hawking `one-boundary' proposal
is incomplete is that conformal modes make the Einstein-Hilbert action
unbounded below, so the path integral seems infinite even without the
ultraviolet divergence \cite{GHP}. If the analogue of histories in
string/M theory that can be well approximated by low-curvature
geometries have actions that are similar to their general-relativistic
approximations, then the string/M theory action would also be unbounded
below and apparently exhibit the same infrared divergences as the
Einstein-Hilbert action for general relativity.  There might be a
uniquely preferred way to get a finite answer by a suitable restriction
of the path integral, but it is not yet clear what that might be.

	A third technical problem with the Hartle-Hawking path integral
is that one is supposed to sum over all four-dimensional geometries,
but the sum over topologies is not computable, since there is no
algorithm for deciding whether two four-dimensional manifolds have the
same topology.  This might conceivably be a problem that it more
amenable in string/M theory, since it seems to allow generalizations of
manifolds, such as orbifolds, and the generalizations may be easier to
sum over than the topologies of manifolds.

	A fourth problem that is likely to plague any proposal for the
quantum state of the cosmos is that even if the path integral could be
uniquely defined in a computable way, it would in practice be very
difficult to compute.  Thus one might be able to deduce only certain
approximate features of the universe from such a path integral.

	One can avoid many of the problems of the Hartle-Hawking
path-integral, and achieve some partial successes, by taking a
`zero-loop' approximation\cite{Page-Hawking-BD}.

\section{Partial Successes of the Hartle-Hawking \\ Proposal}

	Despite the difficulties of precisely defining and evaluating
the Hartle-Hawking `one-boundary' proposal for the quantum state of
the universe, it has had a certain amount of partial successes in
calculating certain approximate predictions for highly simplified toy
models:

\begin{enumerate}

\item Lorentzian-signature spacetime can emerge in a WKB limit of an
analytic continuation
\cite{H,HH}.

\item The universe can inflate to large size
\cite{H}.

\item Models can predict near-critical energy density
\cite{H,HP}.

\item Models can predict low anisotropies
\cite{HL}.

\item Inhomogeneities start in ground states and so can fit cosmic
microwave background data
\cite{HawHal}.

\item Entropy starts low and grows with time
\cite{Haw,Page85,HLL}.

\end{enumerate}

\section{Susskind's Objection to the Hartle-Hawking \\ Proposal}

	Leonard Susskind \cite{Susspriv,DKS,GKS,Sus03}
has argued that the cosmological constant or quintessence or dark
energy that is the source of the present observations of the cosmic
acceleration \cite{Perl,Riess} would
give a large Euclidean 4-hemisphere as an extremum of the
Hartle-Hawking path integral that would apparently swamp the extremum
from rapid early inflation.  Therefore, to very high probability, the
present universe should be very nearly empty de Sitter spacetime,
which is certainly not what we observe.  

	This argument is a variant of Vilenkin's old objection
\cite{VvsHH} that the no-boundary proposal favors a small amount of
inflation, whereas the tunneling wavefunction favors a large amount. 
Other papers have also attacked the Hartle-Hawking wavefunction
\cite{BC,GrT,BP}.  However, Susskind was the first to impress upon me
the challenge to the Hartle-Hawking no-boundary proposal from the
recent cosmic acceleration.

	Of course, it may be pointed out that most of de Sitter
spacetime would not have observers and so would not be observed at
all, so just the fact that such an unobserved universe dominates the
path integral is not necessarily contrary to what we do observe.  To
make observations, we are restricted to the parts of the universe
which have observers.  One should not just take the bare probabilities
for various configurations (such as empty de Sitter spacetime in
comparison with a spacetime that might arise from a period of rapid
early inflation).  Rather, one should consider conditional
probabilities of what observers would see, conditional upon their
existence \cite{Vil95,SQM,HH2}.

	However, the bare probability of an empty de Sitter spacetime
forming by a large 4-hemisphere extremum of the Hartle-Hawking path
integral dominates so strongly over that of a spacetime with an early
period of rapid inflation that even when one includes the factor of
the tiny conditional probability for an observer to appear by a vacuum
fluctuation in empty de Sitter, the joint probability for that
fluctuation in de Sitter dominates over the probability to form an
inflationary universe and thereafter observers by the usual
evolutionary means.  Therefore, the argument goes, almost all
observers will be formed by fluctuations in nearly empty de Sitter,
rather than by the processes that we think occurred in our apparently
inflationary universe.

	The problem then is that almost all of these
fluctuation-observers will not see any significant ordered structures
around them, such as the ordered large-scale universe we observe.
Thus our actual observations would be highly atypical in this
no-boundary wavefunction, counting as strong observational evidence
against this theory (if the calculation of these probabilities has
indeed been done correctly).  As Dyson, Kleban, and Susskind put it
in a more general challenge to theories with a cosmological constant
\cite{DKS}, ``The danger is that there are too many possibilities
which are anthropically acceptable, but not like our universe.''  See
\cite{decay,BF} for further descriptions of this general problem.

	The general nature of this objection was forcefully expressed
by Eddington 75 years ago \cite{Edd}:  ``The {\it crude} assertion
would be that (unless we admit something which is not chance in the
architecture of the universe) it is practically certain that at any
assigned date the universe will be almost in the state of maximum
disorganization.  The {\it amended} assertion is that (unless we admit
something which is not chance in the architecture of the universe) it
is practically certain that a universe containing mathematical
physicists will at any assigned date be in the state of maximum
disorganization which is not inconsistent with the existence of such
creatures.  I think it is quite clear that neither the original nor
the amended version applies.  We are thus driven to admit anti-chance;
and apparently the best thing we can do with it is to sweep it up into
a heap at the beginning of time.''

	In Eddington's language, Susskind's challenge is that the
Hartle-Hawking no-boundary proposal seems to lead to pure chance (the
high-entropy nearly-empty de Sitter spacetime), whereas to meet the
challenge, we need to show instead that somehow in the very early
universe (near, if not at, the ``beginning of time'') it actually
leads to anti-chance, something far from a maximal entropy state.

	For further details of Susskind's challenge, see my recent
account \cite{HH-challenge}.

	This research was supported in part by the Natural Sciences and
Engineering Research Council of Canada.

\eject

\vfill

\end{document}